\documentclass[%
reprint,
amsmath,amssymb,
longbibliography,
pre,
]{revtex4-1}

\usepackage{graphicx}
\usepackage{epstopdf}
\usepackage{dcolumn}
\usepackage{bm}
\usepackage{epstopdf}
\usepackage{xcolor}
\usepackage{sistyle}
\SIthousandsep{,}
\usepackage{amsmath}
\usepackage{float}
\usepackage{color}

\begin{document}
	

\title{Elucidating distinct ion channel populations on the surface of hippocampal neurons via single-particle tracking recurrence analysis}

\author{Grzegorz Sikora$^{1,2}$, Agnieszka Wy\l{}oma\'{n}ska$^1$ and Janusz Gajda$^1$}
\author{Laura Sol\'{e}$^{3}$, Elizabeth J. Akin$^{3}$, Michael M. Tamkun$^{3,4}$}
\author{Diego Krapf$^{ 2,5}$}
\affiliation{$^1$Faculty of Pure and Applied Mathematics, Hugo Steinhaus Center,
Wroc\l{}aw University of Science and Technology, 50-370 Wroc\l{}aw, Poland}
\affiliation{$^2$Department of Electrical and Computer Engineering, Colorado State University, Fort Collins, CO 80523, USA}
\affiliation{$^3$Department of Biomedical Sciences, Colorado State University, Fort Collins, CO 80523, USA}
\affiliation{$^4$Department of Biochemistry and Molecular Biology, Colorado State University, Fort Collins, CO 80523, USA}
\affiliation{$^5$School of Biomedical Engineering, Colorado State University, Fort Collins, CO 80523, USA}
\email{krapf@engr.colostate.edu}

\date{\today}

\begin{abstract}
	
Protein and lipid nanodomains are prevalent on the surface of mammalian cells. In particular, it has been recently recognized that ion channels assemble into surface nanoclusters in the soma of cultured neurons. However, the interactions of these molecules with surface nanodomains display a considerable degree of heterogeneity. Here, we investigate this heterogeneity and develop statistical tools based on the recurrence of individual trajectories to identify subpopulations within ion channels in the neuronal surface. We specifically study the dynamics of the K$^+$ channel Kv1.4 and the Na$^+$ channel Nav1.6 on the surface of cultured hippocampal neurons at the single-molecule level. We find that both these molecules are expressed in two different forms with distinct kinetics with regards to surface interactions, emphasizing the complex proteomic landscape of the neuronal surface. Further, the tools presented in this work provide new methods for the analysis of membrane nanodomains, transient confinement, and identification of populations within single-particle trajectories.       

\end{abstract}


\maketitle 

\section{INTRODUCTION}	

One of the most striking features of mammalian cells lies in their ability to perform extremely intricate functions with a limited number of protein-coding genes. This number is much smaller than originally estimated \cite{pennisi2012encode,ezkurdia2014multiple}. For example, human and mouse genomes have merely 19,817  and 21,968  protein-coding genes (GENCODE 26 and GENCODE M13 \cite{harrow2012gencode}). In order to reach the diversity and complexity required by cells in any mammal, genes can produce multiple protein forms, which can further be chemically modified at several locations. As a consequence, cells can employ the same protein for remarkably different functions. A particular type of proteins that exhibit exceptional diversity are integral membrane proteins, such as receptors and ion channels. It is estimated that approximately 26\% of the human protein-coding genes code for membrane proteins \cite{fagerberg2010prediction}. 

Biological systems are often characterized by both static \cite{xue1995differences} and dynamic \cite{zwanzig1992dynamical} heterogeneities. However, such disorder cannot be usually probed by ensemble-averaged measurements. On the other hand, single-molecule techniques are ideal for observing functional heterogeneities and to extract information on the distribution of molecular properties. Single-molecule experiments have provided information on functional heterogeneities in enzymatic turnover \cite{lu1998single}, RNA folding \cite{solomatin2010multiple}, Holliday junctions \cite{hyeon2012hidden}, and helicase activity \cite{liu2013dna}, to name a few examples. As single-molecule techniques advance, it is becoming clear that functional heterogeneity is ubiquitous in the complex realm of biological systems \cite{ha2001single,moffitt2008recent,hinczewski2016directly}. 

In the plasma membrane functional heterogeneities can be employed to exploit the same protein in multiple cellular functions or to regulate physiological processes by altering intermolecular interactions. For example, besides regulating action potential waveform in neurons, the ion channel Kv2.1 has a non-traditional structural role by which it induces endoplasmic reticulum/plasma membrane contact sites \cite{fox2015induction} and alters membrane protein trafficking \cite{deutsch2012kv2}. Identifying heterogeneities and quantifying the distribution of molecular properties are important steps in cell biology. Single-particle tracking provides unique advantages for the investigation of the dynamics of individual molecules \cite{metzler2014review,krapf2015reviewCTM,manzo2015review}. However, observing heterogeneous dynamics can be challenging due to the inherent thermal fluctuations and experimental noise \cite{ott2013revealing}. Some types of heterogeneous dynamics that have been recognized in trajectories in the plasma membrane include hop-diffusion between actin-delimited membrane compartments \cite{ritchie2003fence,andrews2008actin,sadegh2017plasma}, confinement in nanoscale membrane domains \cite{dahan2003diffusion,garcia2014nanoclustering,AkinBiophys}, and transient tethering to intracellular scaffolds \cite{choquet2013dynamic,weigel2013pnas}. Thus, tools that allow both to identify heterogeneous dynamics in single-particle trajectories and to distinguish particle-to-particle variations in terms of their dynamics, are necessary. Different methods have been developed to identify transition points within intermittent trajectories. For example, a system-level maximum-likelihood method has been employed to identify periods of confined motion within trajectories exhibiting Gaussian diffusion \cite{koo2016systems}. This method is very effective when dealing with Gaussian-based models. Alternatively, universal model-free methods enable the identification of change points in an individual trajectory by considering a local functional that transforms the trajectory into a new time series. This new time series can then be used to characterize intermittent behavior \cite{lanoiselee2017unraveling,wagner2017classification}. Examples of local functionals that have been employed include the diffusivity \cite{persson2013extracting}, convex hull \cite{lanoiselee2017unraveling}, anomalous exponent of the local MSD \cite{weron2017switching}, and directional changes \cite{katrukha2017probing,sadegh2017plasma}. In particular, the local MSD exponent and the convex hull have been used to detect confinement zones. The advantage of local functional methods lies in the fact that they can be applied without prior knowledge of the  model.

In this paper we study the heterogeneous dynamics of two voltage-gated ion channels in the somatic plasma membrane of hippocampal neurons, the K$^+$ channel Kv1.4 and Na$^+$ channel Nav1.6. These channels are observed to be transiently confined in nanoscale domains, but while some molecules remain confined for minutes, others escape in less than 1 s. We introduce a local functional method based on recurrence analysis to identify regions of confinement in the path. Then, we classify trajectories employing a three-step protocol. First, a regime variance test quantifies heterogeneity in particle dynamics. Second, a silhouette analysis is used to identify the exact number of trajectory classes. And third, a $k$-means algorithm is used to set thresholds and separate trajectories into different classes. We find that there are two different classes of trajectories for both Kv1.4 and Nav1.6. These classes of trajectories have very different residence times within the confined domains. While populations that exhibit weak interactions have sojourn times with exponential tails, the populations with strong interactions appear to have heavy tails. These results highlight the complexity of the neuronal surface and provide tools for the study of static and dynamic heterogeneities in the plasma membrane.

\section{MATERIALS AND METHODS}

\subsection{Cell culture, transfection, and labeling} Rat hippocampal neurons were cultured and imaged in glass-bottomed plates as previously described  \cite{akin2015preferential,AkinBiophys}. Animals were used according to protocols approved by the Institutional Animal Care and Use Committee of Colorado State University (Animal Welfare Assurance Number A3572-01).  
Nav1.6 and Kv1.4 constructs were each modified to contain an extracellular biotin acceptor domain (BAD) in an extracellular loop. These constructs (Nav1.6-BAD and Kv1.4-BAD) were previously functionally validated \cite{akin2015preferential}. Neuronal transfections were performed after 6 days in culture for Nav1.6 and 7 days in culture for Kv1.4, using Lipofectamine 2000 (LifeTechnologies, Grand Island, NY). Cells were co-transfected with 1 $\mu$g of either Kv1.4-BAD or Nav1.6-BAD and 1 $\mu$g pSec-BirA (bacterial biotin ligase) to biotinylate the channel.
Labeling of the surface channel was performed before imaging at DIV10. Neurons were rinsed with neuronal imaging saline (NIS), to remove the Neurobasal media. Cells were incubated for 10 min with streptavidin-conjugated CF640R (Biotium, Hayward, CA) diluted 1:1000 in NIS. Streptavidin-CF64R labeling was done at 37$^{\circ}$C in the presence of 1\% bovine serum albumin (cat. A0281, Sigma, St Louis, MO). Excess label was removed by rinsing with neuronal imaging saline.

\subsection{Imaging}
Total internal reflection fluorescence (TIRF) images were acquired at 20 frames per second. 
Before TIRF imaging, differential interference contrast (DIC) and wide-
field fluorescence imaging were used to distinguish transfected neurons
from the relatively flat glia. Neurons were readily identified based on the
characteristic soma morphology and localization of Nav1.6 to the axon
initial segment. All imaging was performed at 37$^{\circ}$C using 
objective and stage heaters.

\subsection{Image processing and single-molecule tracking}
Images were background subtracted and filtered using a Gaussian kernel with a standard deviation
of 0.6 pixels in ImageJ. Tracking of individual fluorophores was then
performed in MATLAB using the U-track automated algorithm \cite{jaqaman2008U-track}. 
Manual inspection confirmed accurate single-molecule detection
and tracking.

\subsection{Identification of transient confinement periods}
In order to identify periods of transient confinement within individual trajectories we developed an algorithm based on trajectory recurrence analysis where we evaluate the total number of visits to the current site \cite{Supplemental}. When a particle is confined within a nanoscale domain, it moves in small area unavoidably visiting the same sites multiple times in a short period. In contrast, during free unconfined motion, the random walk is less compact and its exploration region in the same time is wider. In the recurrence analysis algorithm, at each particle position we calculated the distance to the subsequent point and constructed a circle with diameter equal to this distance, centered midway between the two consecutive points. Next, the number of times the walker position lies within circle, $V'_j$, is calculated. Thus, $V'_j$ denotes the number of visits to site $j$, where $j=1,2,...,N-1$ and $N$ is number of data points in the trajectory. The method by which $V_j$ is found is illustrated in Figs. \ref{fig0}A and B for two simulated trajectories. To improve the algorithm reliability and enhance the differences between confinement and free states, we first segment the data into disjoint windows of size $n=3$ and then sum over the three consecutive $V'_j$ values within each window. For example, $V_1=V_2=V_3=V'_1+V'_2+V'_3$. The identification of states is performed according to $V_j$ remaining either above or below a given threshold ($V_\textrm{th}$). The threshold is selected taking under consideration the behavior of the analyzed data and can vary for diverse data sets. In our experimental data, the threshold was selected to be $V_\textrm{th}=11$; and in synthetic data $V_\textrm{th}=6$. The procedure for threshold selection is further detailed in  Section III.

The statistic $V_j$ is susceptible to statistical noise. Namely, there is a finite, albeit small, probability that $V_j$ crosses the threshold in a single window even though there is no real change of behavior in the data. Nevertheless, the probability that such events take place in two consecutive windows is much smaller. Thus we eliminate most false-positive point changes by considering the dwell time within each state. If a particle crosses the threshold but the dwell time within the new state is only a single time window, i.e., three points, the time series is considered to remain within the same state. 

\subsection{Determination of number of classes among particles}
We determined the number of different trajectory classes based on the time spent in the confined state. After segmentation of trajectories into free and confined states, we calculated the total fraction of time $\phi$ each trajectory resides within the confined state. For a trajectory $i$ with $m$ confined sojourn times $\tau_{ik}$, the fraction of time in the confined state is 
\begin{eqnarray}\label{Ck}
\phi_i=\frac{1}{T_i}\sum_{k=1}^m \tau_{ik},
\end{eqnarray}
where $T_i$ is the observation time. Then, we evaluated if there exist at least two types of trajectories by employing the regime variance test described in Ref. \cite{gajda2013regime}. Briefly, given $M$ trajectories with $\phi_1,\phi_2,...,\phi_M$ fractions of time, we first visually examine trajectory-to-trajectory fluctuations by constructing the successive summation of $\phi^2$,   
\begin{eqnarray}\label{Ck}
C_k=\sum_{i=1}^k\phi_i^2,~~ k=1,2,...,M.
\end{eqnarray}
If the fractions $\phi_i$ correspond to a single type of trajectories, then $C_k$ is a linear function with respect to $k$, otherwise a piecewise linear behavior with different slopes indicates there are at least two different regimes. Note that the values $\phi_i$ do not need to be ordered and the $C_k$ statistics is non-decreasing. The regime variance test was shown in Ref. \cite{gajda2013regime} to be effective when dealing both with Gaussian and L\'{e}vy-stable random variables.

The null hypothesis of the regime variance test corresponds to the case with a single regime. To test this hypothesis, first the most likely switching point $k'$ is found on the basis of the $C_k$ statistic. To find the switching point $k'$, we fit two regression lines to the arrays $C_1,\cdots,C_k$ and $C_{k+1},\cdots,C_M$ and calculate the squared sums of residuals for both lines. The switching point $k'$ is obtained by minimizing the total squared sum of residuals. Next, the data $\{ \phi_i \}$ are divided into two arrays: $\phi_1,\cdots,\phi_{k'}$ and $\phi_{k'+1},\cdots,\phi_M$. Then, given a desired confidence level $\gamma$, the quantiles $(1-\gamma)/2$ and $\gamma/2$ of the squared data $\phi_i^2$ for the first array ($i=1,\cdots,k'$) are calculated. Here, for the sake of simplicity, we assume the variance of the first array is smaller than the second one; otherwise the quantiles are computed for the second array. The core of the regime variance test is the number of observations $B$ from the data $\phi_j^2$ in the second array ($j=k'+1,\cdots,M$) that fall into the constructed quantiles interval. The null hypothesis of both regions having the same distribution implies that $B$ has binomial distribution 
$P(B=k')=\binom{M-k'}{B}\hat{p}^B(1-\hat{p})^{M-k'-B}$, where $\hat{p}=\gamma$. Therefore the $p$-value of the test is equal to the cumulative distribution function of this distribution evaluated at $B$. A large $p$-value of the test (greater than the confidence level $\gamma$) indicates the null hypothesis is not rejected.

After confirming the existence of at least two types of trajectories with respect to sojourn times in the confined states, we determined the number of classes on the basis of the silhouette criterion \cite{s13}. For a fixed number of classes $c$ the silhouette statistic assigns value $s_c(i)$ to the observation $\phi_i$, given by
\begin{equation}
	s_c(i)=\frac{b(\phi_i)-a(\phi_i)}{\textrm{max}\{a(\phi_i),b(\phi_i)\}},~~ i=1,2,...,M,
\label{silhouette}
\end{equation}
where $a(\phi_i)$ is the average distance to all values in the allocated class and $b(\phi_i)$ is the distance to the nearest neighbor class. For each number of classes $c$, all possible divisions into $c$ classes are considered and the optimal division is the one that maximizes the silhouette statistic $s_c$. The silhouette criterion then takes a value 
\begin{equation}
	\textrm{sil}_c=\sum_{i=1}^M s_c(i)/M,
\label{silhouette}
\end{equation}
that varies from $0$ to $1$. The optimal number of classes maximizes $\textrm{sil}_c$.  

\subsection{Classification of trajectories}
In order to classify trajectories according to their fractions of time being in the confined state $\phi_i$, we used a clustering method based on the $k$-means algorithm \cite{Jain} implemented in MATLAB. The $k$-means clustering partitions the set of $M$ observations into $k$ clusters in a way that each observation belongs to the cluster with the nearest mean. The assignment of $\phi_i$ into a cluster is thus based on the minimization of the average Euclidean distance between the points in that cluster and the cluster mean.  This method yields a partitioning of the data space into Voronoi cells. 

\subsection{Statistics}
In the experimental data analysis we compared the distributions of different characteristics corresponding to confinement states for classified trajectories. In order to evaluate if two data sets have the same distribution we used the Kolmogorov-Smirnov (KS) test for two samples \cite{ks}. The KS statistic for two data sets with cumulative distribution functions $F_1(x)$ and $F_2(x)$ is
\begin{eqnarray}
KS=\textrm{sup}_{x}|F_1(x)-F_2(x)|,
\end{eqnarray}
where $\textrm{sup}_{x}$ is the supremum. A large $p$-value of the KS test indicates the $H_0$ hypothesis is not rejected and the two data sets have the same distribution.

\section{Validation of confinement identification method}
In this section we evaluate the effectiveness of the confinement identification method. As the toy model we analyze intermittent fractional Brownian motion (FBM). FBM is a stochastic process driven by stationary Gaussian, but power-law correlated noise \cite{beran1994statistics,mandelbrot1968fractional}. It is one of the classical anomalous diffusion process for which the mean square displacement (MSD) $\langle x^2(t)\rangle=K_\alpha t^{\alpha}$, with generalized diffusion coefficient $K_\alpha$ and anomalous exponent $\alpha$. In terms of the commonly used Hurst exponent, $H=\alpha/2$.  The process is superdiffusive when $\alpha>1$ and subdiffusive when $0<\alpha<1$. As $\alpha$ decreases, the random walk becomes more compact. In particular, when $\alpha$ is close to zero, the FBM resembles confinement in a domain with a small drift. 

\begin{figure}[ht]
	\includegraphics{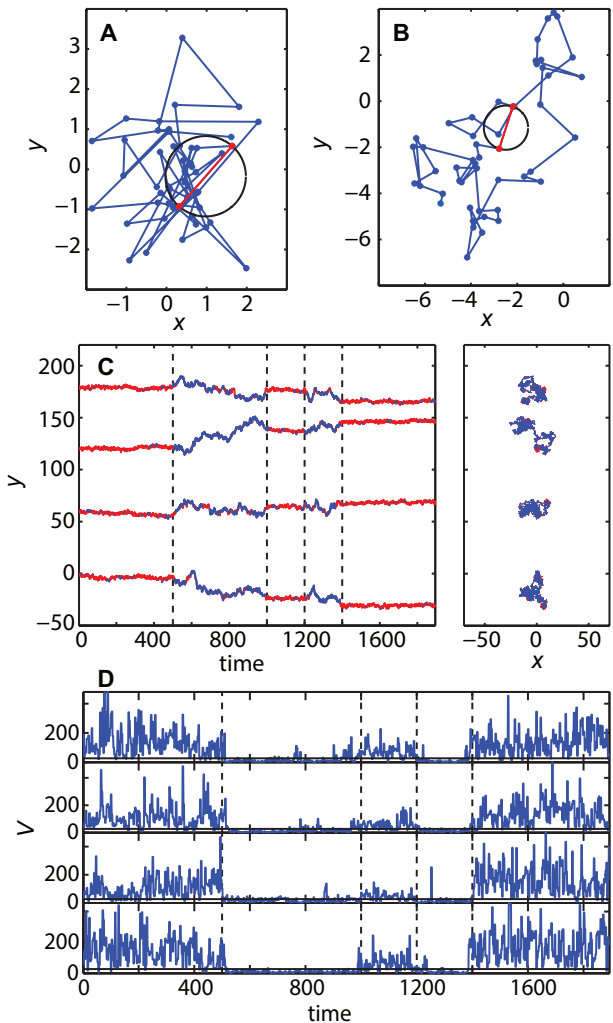}
\caption{Method for identification of confined motion based on number of visits to current site. The method is validated using numerical simulations.
(A) FBM with $\alpha=0.1$, i.e., Hurst exponent $H=0.05$. At each successive points pair such as those indicated by the arrows a circle is drawn as indicated in the figure. In this case there are 17 points (visits) found within the circular area.  
(B) FBM with $\alpha=0.9$. Here the tracer visits the region within the selected circle only once.
(C) $y$-coordinate of two-dimensional intermittent FBM with $\alpha=0.1$ (first, third and fifth segment) and $\alpha=0.9$ (second and fourth segment). The vertical dashed lines show the true switching points. The parts of the trajectories identified as confined (low $\alpha$) are marked in red. The two-dimensional trajectories are presented on the right.
(D) $V_j$ statistic, i.e., number of recurrent visits, as a function of time. Again, the vertical dashed lines show the true switching points. The threshold for finding confined regions in this example is $V_{\textrm{th}}=6$.
}
 	\label{fig0}
\end{figure}

In order to illustrate the confinement recurrence analysis, we present two short FBM trajectories ($N=50$ points) with $\alpha=0.1$ and $\alpha=0.9$ in Fig. \ref{fig0}A and B, respectively. Given that a FBM with $\alpha=0.1$ is a very compact random walk, it resembles motion in a confined domain. FBM with $\alpha=0.9$ is a good model for unconfined subdiffusion. We expect the number of visits $V_j$ to have larger values in the regions with $\alpha=0.1$. Thus, the number of visits provide a metric to segment the trajectory according to its recurrence. For both cases two consecutive observations are marked in Fig. \ref{fig0}A and B. The constructed circle in the trajectory with $\alpha=0.1$ encloses 17 points and the circle in the trajectory with $\alpha=0.9$ encloses only one point.

We analyze an intermittent FBM where the anomalous diffusion exponent alternates between $\alpha=0.1$ and $\alpha=0.9$. For simplicity the random walk is defined as a renewal process where the process correlations are reset when $\alpha$ changes. We simulate the intermittent FBM with five segments of different lengths. The first, third and fifth segments correspond to $\alpha=0.1$ while the second and fourth to $\alpha=0.9$. Fig. \ref{fig0}C shows four simulated intermittent FBM realizations together with the results of the recurrence analysis method. The parts of the trajectories identified as confined motion are marked in red. The recurrence analysis takes under consideration two-dimensional trajectories but we present one-dimensional time series for clarity. The two dimensional trajectories are shown to the right of the time traces. The vertical dashed lines correspond to the switching points between the two FBMs. The time series of the number of visits $V_j$ for these four simulated trajectories are presented in Fig. \ref{fig0} D. Again, the vertical dashed lines correspond to the true switching points between the two regimes of intermittent FBMs. 
As seen in Fig. \ref{fig0}D, the time series $V_j$ remains for long times at low values that correspond to free diffusion and high $V_j$ values corresponding to confined motion. Thus it is possible to discriminate between different phases of motion by employing a threshold on the $V_j$ time series. The choice of the threshold value for segmentation of the trajectories depends on the character of the data but it can be effectively chosen by visual inspection. Here, we chose a threshold $V_{\textrm{th}}=6$. However, as can be seen in Fig. \ref{fig0}D the method is prone to statistical noise: in the free regions we observe falsely identified short periods of confinement and vice versa. Therefore, there is need to correct the method in order to overcome the falsely identified regions with short dwell times. As explained in the methods, this correction is introduced by eliminating all transitions where the dwell time is a single time window, that is three points.

\section{RESULTS}

\subsection{Detection of Kv1.4 transient confinement}
\begin{figure}
  \includegraphics{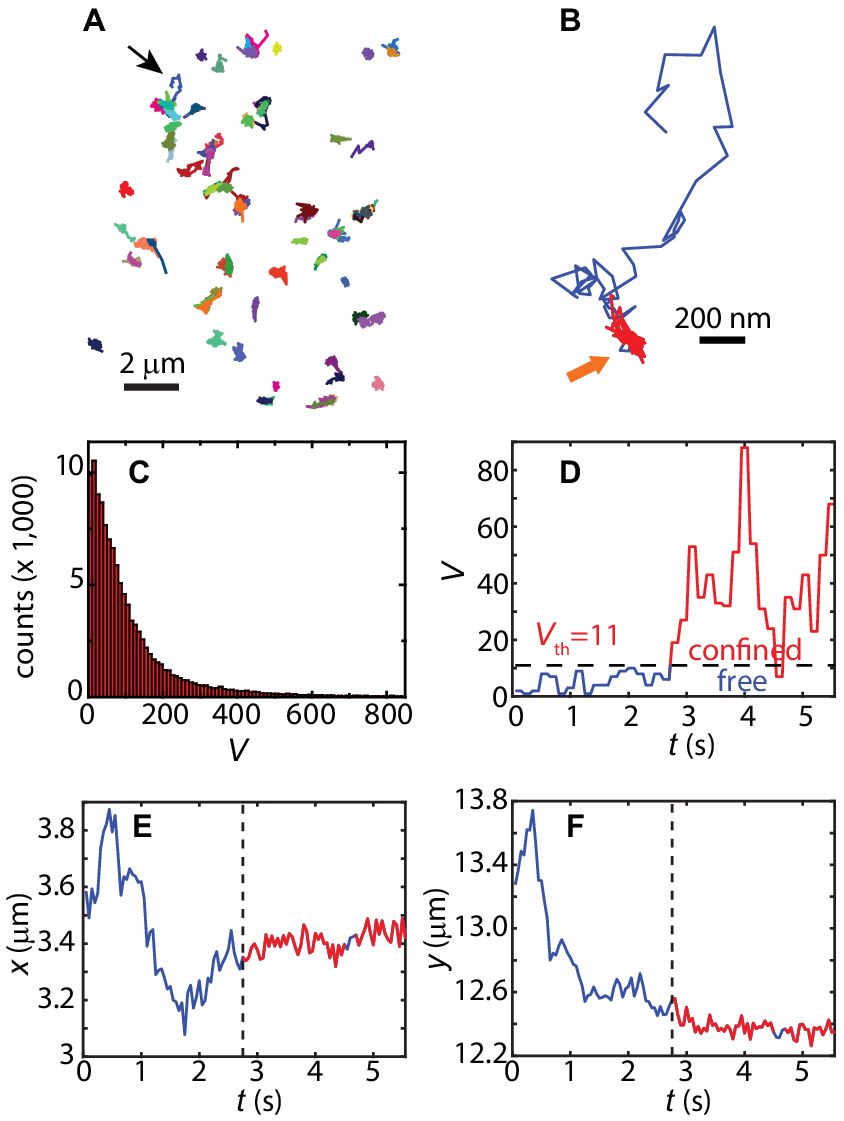}
 	\caption{Segmentation of Kv1.4 trajectories.
	(A) 92 individual Kv1.4 trajectories in one hippocampal neuron.
	(B) Zoom of a trajectory where the periods of confinement are shown in red.
	(C) Histogram of the number of visits to current site, $V_j$ (117,195 sites; 649 trajectories). As explained in the text, the sites are determined by the exploration in 150 ms. This metric is used to evaluate how compact the random walk is and to identify regions of confinement. 
	(D) Time series of number of visits $V_j$ for the trajectory in panel B. When the number of visits is above the threshold $V_{\textrm{th}}=11$, the particle is considered to be in the confined state. At 4.6 s, $V_j$ drops below the threshold during three points in the time series. However, because the dwell time is not longer than three points, i.e., a single time window, this region is considered to remain in the confined state. This segment is marked in blue and indicated by an arrow in panel B.  
	(E-F) Time series of localization along $x$ and $y$ for the trajectory shown in panel B. The regions that are detected to be confined are shown in red and the identified switching point marked by dashed vertical line.}
 	\label{fig:Tracks14}
\end{figure}

We have imaged hippocampal neurons expressing Kv1.4-CF640R and Nav1.6-CF640R and tracked their motion on the somatic surface. Figure~\ref{fig:Tracks14}A shows 92 Kv1.4-CF640R trajectories obtained in a typical cell. The trajectories are highly heterogeneous with some trajectories being very compact while others explore large regions. However, this heterogeneity does not appear to be related to the location of the molecules within the cell. Figure~\ref{fig:Tracks14}B shows a zoom on the trajectory indicated by an arrow. As we have previously reported \cite{weron2017switching}, Kv1.4 ion channels exhibit intermittent behavior with periods of confinement and periods of free diffusion.

We employ recurrence analysis based on the number $V_j$ of visits to site $j$, to segment the trajectory according to being in either a confined or free state. Figure~\ref{fig:Tracks14}C shows an histogram of the number of visits $V_j$ at each site as defined in our algorithm for identification of confinement. The trajectories are next segmented using a threshold $V_\textrm{th}=11$. Figure \ref{fig:Tracks14}D shows the time series $V_j$ of the trajectory shown in panel B. We add the number of visits within three consecutive circles and thus our temporal resolution is 150 ms in this analysis. The confined regions are found as the periods with $V_j\ge V_\textrm{th}$ and are colored in red in Figs. \ref{fig:Tracks14}B and D. The $x(t)$ and $y(t)$ time series of the same trajectory are shown in Figs. \ref{fig:Tracks14}E and F, also with the confined regions colored in red.

\subsection{Kv1.4 are classified according to their surface interactions} 

We have observed that Kv1.4 channels exhibit periods of transient confinement and the instantaneous state of the protein can be determined by recurrence analysis. Further, visual examination of the trajectories in Fig. \ref{fig:Tracks14}A suggests the data are markedly heterogeneous. Thus we study whether there are more than one class of particles using the regime variance method according to the fraction of time $\phi$ that each particle spends in the confined state. From a physiological perspective such different types of molecules could be the result of post-translational modifications that alter molecular interactions. Figure \ref{fig:silhouette}A shows an histogram of the fractions of time spent in the confined state where the counts indicate number of trajectories. These fractions of time vary from $\phi=0.02$ up to $\phi=1$. Figure \ref{fig:silhouette}B shows the regime variance statistic $C_k$ vs. trajectory number $k$ (Eq.~(\ref{Ck})). Using this metric with a confidence level $\gamma=0.05$, we find that there are at least two distinct classes of trajectories in the Kv1.4 data ($p=10^{-3}$). Using the silhouette criterion, we find that there are two classes of trajectories as $\textrm{sil}_c$ is maximized by $c=2$ ({$\textrm{sil}_2=0.9$ and $\textrm{sil}_3=0.8$). As a simple control of the regime variance test, we apply it to the simulated trajectory set of intermittent FBM that was presented in Fig. \ref{fig0}. The regime variance test for the simulated trajectories does not reject the hypothesis of a single class ($p=0.29$, inset of Fig. \ref{fig:silhouette}B).  

Kv1.4 trajectories are classified according to their fraction of time in the confined regime $\phi$. The $k$-means algorithm yields class division according to $\phi<0.69$. Figures \ref{fig:silhouette}C and D show examples of trajectories in each of the classes, where the confined states are marked in red. To characterize the differences in the behavior of particles belonging to each class we study the distributions of residence times within the confined state. Figure \ref{fig:silhouette}E shows the complementary cumulative distribution function (CCDF) of the residence time, i.e., $P[T>t]$ for particles in each of the classes. The Kolmogorov-Smirnov two-sample test rejects the null hypothesis of the same distribution of the residence time for two classes with $p=10^{-108}$. For the trajectories with $\phi<0.69$ we find that the sojourn times have an exponential distribution tail with a characteristic decay time $\tau=0.43\pm 0.05$ s. 
  
\begin{figure}
 	\includegraphics{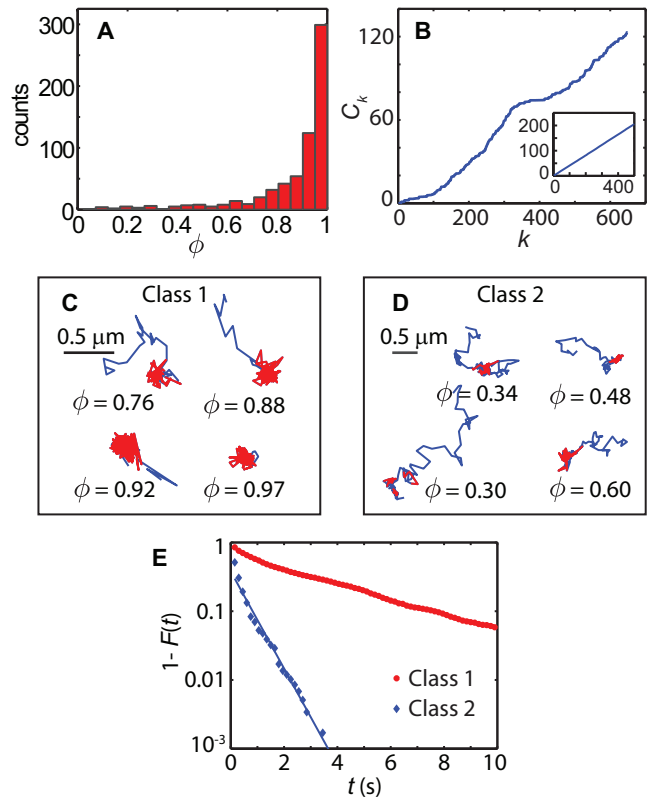}
 	\caption{Determination of number of distinct classes among Kv1.4 trajectories (649 trajectories, 9 cells). 
(A) Histogram of fraction of the observation time that a particle spends in the confined state. 
(B) Regime variance test statistic $C_k=\sum_{i=1}^k\phi_i^2$.  The regime variance test rejects the hypothesis of a single regime with $p=10^{-3}$. The inset shows the regime variance test applied to 500 realizations of numerical simulations of the type shown in Fig. \ref{fig0}. In these simulations, the test does not reject the hypothesis of a single class of trajectories ($p=0.29$).  
(C-D) Examples of Kv1.4 trajectories in the two distinct classes. The fraction $\phi$ is given for each example.
(E) Complementary cumulative distribution function of the residence times in confined states for each of the two classes (class 1: 1,659 sojourn times, 575 trajectories; class 2: 575 sojourn times, 74 trajectories).}
 	\label{fig:silhouette}
\end{figure}


\subsection{Characterization of Kv1.4 confining domains} 

The confining domains were found from the periods within the trajectories that particles exhibit confined motion. However, only regions where the particle remains confined for at least 10 frames were analyzed. In the cases that the particle was confined for more than 20 frames, only the first 20 points are employed in the analysis to avoid any potential problems related to drift of the confining domain. Radii of gyration were found along the major and minor principal axes and the domain was approximated as an ellipse with these major and minor semi-axes, respectively, as shown in the inset of Fig. \ref{fig:Domains}A. The radius of gyration along the $u$ direction is defined for a trajectory of $N$ points as $R_u^2=\sum_{k}^N(u_{k}-\overline{u})^2$, where $u_k$ is the distance of the $k^\textrm{th}$ point to the corresponding principal axis. Figures \ref{fig:Domains} A and B show the cumulative distribution function and box plots of the elliptical area of the confining domains for both classes of trajectories. Figure \ref{fig:Domains} C and D show the characterization of the same domains in terms of the major semi-axis. The majority of domain sizes are much larger that the particle localization uncertainty. These data show that even though the interactions of distinct pools of Kv1.4 with these domains are different according to their sojourn times, the confining domains for both classes share similar morphological characteristics.  

\begin{figure}
 	\includegraphics{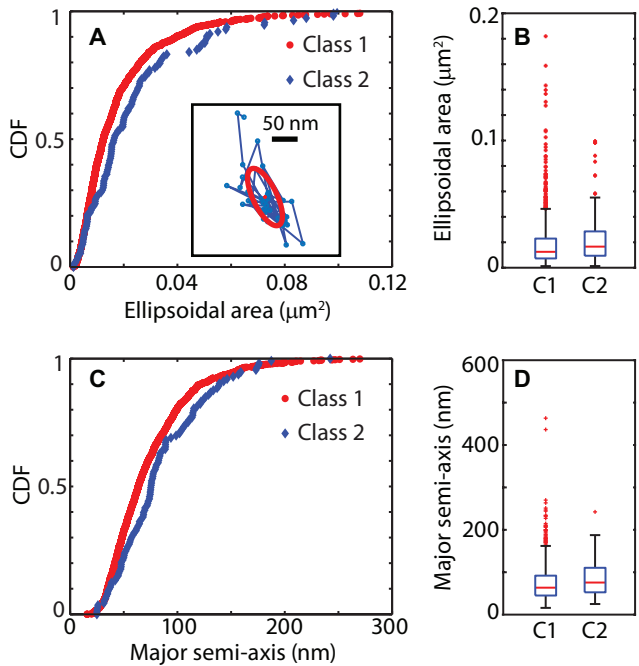}
 	\caption{
	Confining regions are the same for the particles in both classes.
(A) Cumulative distribution function and (B) box plots for the confining regions area for particles in both classes. Box plots show 5-95\% quantiles as whiskers together with quartiles and median (Class 1: 1,165 domains; Class 2: 113 domains). The inset in panel A shows the characterization of the confining domain in a sample trajectory using the radii of gyration along the principal axes. The first 20 points of the confining domain are employed to find the radii of gyration. The obtained confining ellipse is shown together with the principal axes. 
(C) Cumulative distribution function and (D) box plots for the confining major semi-axes.
}
 	\label{fig:Domains}
\end{figure}


\subsection{Characterization of Nav1.6 motion} 

\begin{figure}
 	\includegraphics{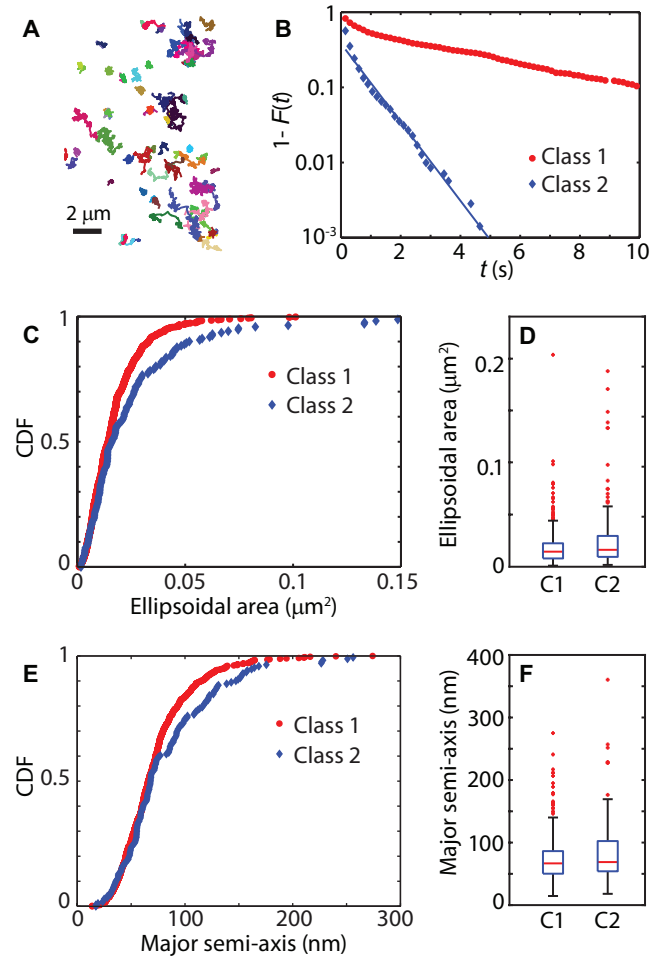}
 	\caption{
	Nav1.6 channels in the soma of hippocampal neurons (386 trajectories, 4 cells).
(A)	Trajectories of Nav1.6 in one example cell (79 trajectories).
(B) Residence times in confining regions for the two classes found using the silhouette method (Class 1: 915 times, 307 trajectories; class 2: 706 times, 79 trajectories).
(C) Cumulative distribution function (CDF) and (D) box plots of the confining regions area for particles in both classes. Box plots show 5-95\% quantiles as whiskers together with quartiles and median (Class 1: 605 domains; class 2: 171 domains). 
(E) CDF and (F) box plots of the confining major semi-axes.
}
 	\label{fig:Nav}
\end{figure}

Nav1.6 ion channels were previously also found to exhibit periods of transient confinement \cite{AkinBiophys}. Nav1.6 trajectories in a representative cell are shown in Fig. \ref{fig:Nav}A. Regime variance test also shows that there exist at least two classes of Nav1.6 trajectories and the silhouette criterion indicates the number of classes equals two ($\textrm{sil}_2=0.89$ and $\textrm{sil}_3=0.75$). Application of a $k$-means algorithm yields a fraction of times threshold $\phi=0.74$ for classifying trajectories. The distributions of sojourn times in the confined state are shown Fig. \ref{fig:Nav}B. Again, as seen for Kv1.4 channels, the distributions of times in the two states are markedly different. The sojourn times in the class with $\phi<0.74$ are exponentially distributed with characteristic decay time $\tau=0.93 \pm 0.04$ s.

The characterization of confining domains for Nav1.6 according to the radii of gyration along the principal axes is shown in Figs.  \ref{fig:Nav} C-F. The Kolomogorov-Smirnov two-sample test rejects the null hypothesis of the same distribution of confinement sizes for two classes with $p=0.007$ but the characteristics of both populations are similar. 


\section{DISCUSSION AND CONCLUSIONS}
The dynamics of membrane proteins is often characterized by a high degree of heterogeneity. In general, these fluctuations can arise from two very different mechanisms. In the first situation, proteins perform a random walk in a heterogeneous landscape while, in the second, proteins undergo post-translational modifications so that they interact in substantially different ways with the same complexes. The first situation has been studied both experimentally and theoretically. Heterogeneous diffusion landscapes can yield intriguing results that involve population splitting and non-ergodicity \cite{cherstvy2013population,manzo2015PRX}. Besides analysis of individual trajectories, the diffusion landscape of membrane proteins has been studied using single-particle tracking photoactivated localization microscopy (sptPALM) \cite{manley2008high} and universal points-accumulation-for-imaging-in-nanoscale-topography (uPAINT) \cite{giannone2010dynamic}, which yield high-density surface maps. Furthermore, these high-density maps can be accurately evaluated using Bayesian inference tools, which provide information on both diffusion and energy landscapes \cite{masson2014mapping}.

We have previously employed sptPALM in combination with Bayesian inference tools to show that Nav1.6 channels are clustered into nanoscale domains \cite{AkinBiophys}. However, one of the interesting aspects of those observations lies in the fact that Nav channels exhibit a marked heterogeneity in their interaction with the nanoclusters. Therefore we set to study the molecule-to-molecule heterogeneity in the neuronal surface. We raise the question, can we unravel distinct molecule subpopulations according to interactions with nanoclusters? To this end we develop a methodology by which we segment the trajectories into regions of transient confinement and regions of free motion and we identify two distinctive subpopulations both in the Nav1.6 and in the Kv1.4 dynamics. These subpopulations exhibit different types of interactions with their respective membrane nanodomains. In one population the trajectories are mostly in the free state while in the second population, the trajectories exhibit long periods under transient confinement. The size of the confined regions are characterized from the motion of the molecules and it is found these regions have a mean diameter that is ten times the localization accuracy. Therefore the trapping events cannot be considered to be immobilization due to binding as is the case for Kv2.1 channels in HEK cells \cite{weigel2011PNAS,weigel2013pnas}.

The interactions of one of the populations with the confining nanodomains exhibit a `normal' type of statistics with sojourn times that are exponentially distributed. Therefore, the system can be considered to be Markovian, i.e., to have no memory. Surprisingly, the second population exhibits a heavy-tail, non-exponential sojourn-time distribution. This behavior brings up the hypothesis of complex behavior with the possibility of ergodicity breaking and aging in the dynamics of the ion channels. Consistent with these observations, we have recently found that the dynamics of the majority of Kv1.4 and Nav1.6 trajectories in the somatic plasma membrane exhibit non-ergodic dynamics according to dynamical functional tests \cite{weron2017switching}.

The tools developed in this work can be employed in the study of membrane nanodomains, which are widespread among mammalian cells. Further these domains can play important physiological roles. 
In B and T lymphocytes, reorganization of signaling nanodomains leads to cell activation \cite{pizzo2004lipid}. In neurons, nanoclustering of membrane proteins has key functions in synaptic transmission \cite{dani2010superresolution}. We apply four time-series analysis tools to extract specific information on heterogeneous interactions with nanodomains: (i) A recurrence analysis is used to find transitions in the diffusive behavior. This analysis has the advantages of providing high temporal resolution that can be applied to both Markovian and non-Markovian processes. (ii) A regime variance test quantifies the heterogeneity in the sojourn times. (iii) A silhouette algorithm finds the number of different classes according to protein dynamics. (iv) A $k$-means algorithm is used to set thresholds and separate trajectories into different classes. By using the algorithms provided in the Supplemental Materials \cite{Supplemental} we identified and characterized distinct behaviors of the same proteins expressed in the neuronal soma, which had not been distinguished with previous analyses.

\begin{acknowledgments}
This work was supported by the National Science Foundation under Grant 1401432 (to DK), NCN OPUS Grant No. UMO-2016/21/B/ST1/00929 (to AW), NCN Maestro Grant No. 2012/06/A/ST1/00258 (to JG),​ and the National Institutes of Health grant RO1NS085142 (to MMT). 
\end{acknowledgments}

\providecommand{\noopsort}[1]{}\providecommand{\singleletter}[1]{#1}%

\end{document}